\documentclass[%
 reprint,
superscriptaddress,
 amsmath,amssymb,
 aps,
]{revtex4-2}

\usepackage[english]{babel}

\usepackage{amsmath}
\usepackage{graphicx}
\usepackage[autostyle]{csquotes}
\usepackage[colorlinks=true, allcolors=blue]{hyperref}
\usepackage{multirow}
\usepackage{mathrsfs}
\usepackage{upgreek}
\usepackage{lineno}
\usepackage{epstopdf}
\usepackage{subcaption}
\usepackage{ulem}

\begin{document}


\title{Phonon-mediated stabilization of first and second modes\\ in hypersonic boundary-layer flows}

\author{Christoph Brehm}
\email[]{cbrehm1@umd.edu}
\affiliation{Department of Aerospace Engineering, University of Maryland, College Park, Maryland 20740, USA}
\author{Connor W. Klauss}
\email[]{cwklauss@umd.edu}
\affiliation{Department of Aerospace Engineering, University of Maryland, College Park, Maryland 20740, USA}
\author{Mahmoud I. Hussein}
\email[]{mih@colorado.edu}
\affiliation{Smead Department of Aerospace Engineering Sciences, University of Colorado Boulder, Boulder, Colorado 80303, USA}
\affiliation{Department of Physics, University of Colorado Boulder, Boulder, Colorado 80302, USA}

\begin{abstract}
Laminar–to-turbulent transition delay is a key challenge in hypersonic boundary-layer flows. Unstable disturbances$-$most prominently the first and second modes$-$trigger the onset of turbulence and pose a fundamental technological barrier to hypersonic transport. While existing control strategies target the second mode, simultaneous mitigation of the first mode has long appeared physically impossible. A new flow-control concept is introduced in which phase relations between wall pressure and velocity fluctuations are tailored using subsurface phonon engineering to control both modes concurrently. The outcome is substantial drag reduction and alleviation of the extreme thermal loads associated with turbulence.
\end{abstract}

\maketitle
Laminar-to-turbulent boundary-layer transition (BLT) in hypersonic flows causes substantial increases in surface heat transfer and skin-friction drag, posing major challenges for vehicle structural design, material selection, and overall performance \cite{schneider2004hypersonic,hollis2013heating}. The resulting rise in surface temperature places severe constraints on thermal protection systems and structural materials, increasing manufacturing cost but most crucially limiting the operational speed and range of hypersonic vehicles. A recent market study of commercial hypersonic transport under a NASA contract \cite{Bastedo2021HypersonicMarket} suggests that cruise Mach numbers in the range of 4–6 could enable travel across the globe within approximately 2–4 hours. Although numerous technological obstacles remain, including reusable thermal protection systems and efficient propulsion \cite{pollock2024examination}, maintaining laminar flow over a large portion of the vehicle surface would have a transformative effect on the economic viability of high-speed transport \cite{pfenninger1988design,powell1989laminar,woan1991LFC,frohler2025economic}. This prospect has motivated extensive efforts  to prevent or delay laminar-to-turbulent transition over hypersonic vehicle surfaces. However, robust transition control strategies remain elusive due to the presence of multiple instability mechanisms that provide competing pathways to turbulence.

In the targeted Mach number range of approximately 4–6, transition in hypersonic boundary layers is primarily governed by Mack’s first- and second-mode instabilities~\footnote{These instabilities are also referred to, interchangeably, as \textit{disturbances} or \textit{fluctuations}}~\cite{ack1969boundaryJPLRep}.~These two instability mechanisms arise from fundamentally different physical processes \cite{liang2022inviscidenergetics}. A key example highlighting this difference is their opposite response to wall cooling: while wall cooling stabilizes the first mode, it enhances the growth of the second mode, and vice versa~\cite{ack1969boundaryJPLRep}. Another important distinction is that the most amplified first-mode wave is oblique with a non-zero spanwise wavenumber, $\beta\neq0$, whereas second-mode disturbances are two-dimensional, i.e. $\beta$$=$$0$,  and propagate strictly in the base-flow direction. Over the past several decades, extensive research has focused on passive flow-control strategies targeting second-mode stabilization, most prominently through porous-wall treatments~\cite{fedorov2001porous,bres2010porouscoating,bres2013secondmode} and acoustic metasurfaces ~\cite{zhao2018optimalporous,zhao2019impedencenearzero,zhao2021controlreflectedwaves,zhao2022acousticmetasurfacereview}. Although these approaches can attenuate second-mode disturbances, they generally fail to control the first mode---in many cases, stabilization of the second mode is accompanied by simultaneous destabilization of the first mode~\cite{fedorov2001porous, tullio2010porous,wang2012localporous,tritarelli2015feltmetal}. It has also been hypothesized by several renowned researchers \cite{lees1947stability,malik1989prediction,kara2008wallcooling} in the field that stabilization of the first mode through passive wall treatments alone may be impossible without modifying the mean flow, for example through wall cooling. As a result, no passive flow-control strategy has yet been demonstrated to simultaneously stabilize both first- and second-mode instabilities.

\begin{figure}[b]
\centering
\includegraphics[width=\linewidth]{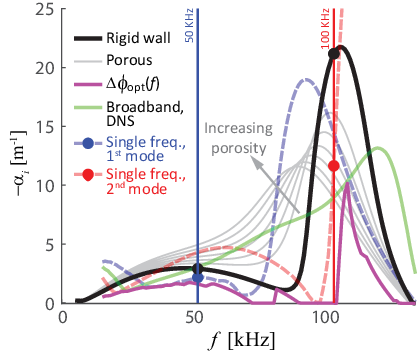}
\caption{Spatial amplification rates, $-\alpha_i$, obtained from LST of Mach 5.35 boundary-layer profiles extracted at $x=0.6$ m for a rigid wall (RW), porous walls with different porosities, and multiple phonon-engineered subsurfaces: `per frequency' optimized $\Delta \phi_{opt}\left(f \right)$ PSub, broadband optimized PSub used in the DNS Fig.~\ref{fig:breakdownDelay}, PSub with $\Delta\phi_{p^\prime_w,v^\prime_w}=75^\circ$ for first-mode stabilization at 50 kHz, and PSub with$\Delta\phi_{p^\prime_w,v^\prime_w}=275^\circ$ for the second-mode stabilization at 100 kHz.}

\label{fig:LSTFMSM}
\end{figure}

In stark departure from existing hypersonic flow control strategies, this work introduces an approach based on local phase control between wall-pressure and velocity fluctuations, enabling simultaneous stabilization of both instability modes. The underlying passive phase engineering, along with accompanying amplitude control, is realized by \textit{subsurface phonon mediation}~\cite{hussein2015flowstab}. Figure \ref{fig:LSTFMSM} provides results from a local stability theory (LST) analysis \cite{Tollmien1929,al2023biorthogonal} of a Mach 5.35 adiabatic boundary-layer profile extracted at $x=0.6 \text{ m}$ downstream of the leading edge for a disturbance frequency of 100 kHz, and more generally across the full spectrum.  Under the locally parallel-flow assumption of LST, disturbances are represented by the spatial traveling-wave ansatz $\boldsymbol{q}^\prime\left( x,y,z,t\right)=[p^\prime,u^\prime_1,u^\prime_2,u^\prime_3,T^\prime]^T =\tilde{\boldsymbol{q}}(y)\exp\left[i(-\alpha x+\omega t+\beta z)\right]$ \cite{ack1969boundaryJPLRep}. The spatial amplification rate (with $-\alpha_i>0$ for unstable waves) shows that conventional porous media will stabilize the second mode but destabilize the first mode; crucially, no porous-wall design achieves stabilization of the first mode. In contrast, our approach targets phase manipulation between wall pressure and velocity fluctuations to intervene with both instability modes at once. Such control utility is realizable using phononic subsurfaces (PSubs)~\cite{hussein2015flowstab}. Phononics is an emerging field with rapidly expanding applications in applied physics and engineering owing to its ability to realize architected structures with precisely tunable frequency-dependent amplitude and phase responses~\cite{hussein2014phononicsreview}. Most importantly, recent advances in fabrication technology for high-temperature lattices~\cite{wang2025additive} are now opening a path for phononics in hypersonic applications. In a recent study, PSubs were explored experimentally for attenuating shock-induced boundary-layer oscillations~\cite{navarro2025stabilization}. This Letter targets the fluctuation production mechanism governing Mack-mode growth. Using controlled-transition direct numerical simulations (DNS), we demonstrate simultaneous stabilization of the first and second modes under monochromatic as well as more realistic broadband disturbance environments. We first derive the theoretical limits of the control strategy considering monochromatic disturbances and demonstrate its effectiveness, and then extend to a more complex phase/amplitude-control intervention targeting broadband disturbances.


\begin{figure}[h]
\centering
\includegraphics[width=\linewidth]{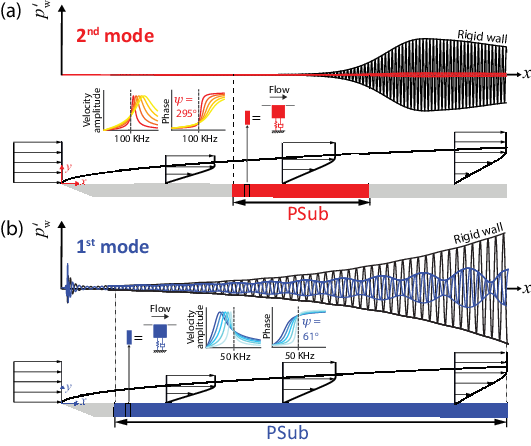}
\caption{Wall-pressure fluctuation signals for rigid walls (black) and with PSub (red/blue). The second-mode is introduced at $f=100 \mathrm{kHz}$ and the first-mode at $f=50 \mathrm{kHz}$. Insets show frequency response functions of the respective PSubs.} 
\label{fig:amplitude-curves}
\end{figure}


 We consider a Mach~5.35 adiabatic-wall boundary layer in which both first- and second-mode instability waves are present, as shown in Fig.~\ref{fig:LSTFMSM}, at a unit Reynolds number of $10\times10^{6}$ m$^{-1}$. The computational domain extends from $x=0.1$ m to $x=1.0$ m. Controlled-transition DNS are performed by introducing disturbances through a volume-forcing term placed inside the boundary layer following the approach of Ref.~\cite{browne2022nonlinear}. As disturbances propagate downstream within the boundary layer, their evolution is tracked through wall-pressure fluctuations as shown in Fig.~\ref{fig:amplitude-curves}. For the rigid wall (RW) case (black line), the 2-D instability waves with disturbance frequencies of $50\,\mathrm{kHz}$ and $100\,\mathrm{kHz}$ associated with the first and second mode instabilities, respectively, grow exponentially as expected for an unstable boundary-layer flow. When a phonon engineered subsurface is inserted, however, the disturbances are strongly attenuated in each case, demonstrating stabilization of both first- and second-mode waves. The wall response was obtained from 2-D fully-coupled fluid–structure interaction simulations in which local spring–mass–damper systems, with natural frequencies of 40 kHz and 120 kHz and corresponding damping ratios of 0.13 and 0.09, respectively, are driven by the instability-wave–induced wall-pressure fluctuations.

The underlying framework requires a prescribed relationship between wall-pressure and wall-normal velocity disturbances such that the phased coupling suppresses instability growth:
\begin{equation}
\tilde{v}_w(x,\omega) = \mathcal{H}(\omega)\tilde{p}_w(x,\omega),
\label{equ:transfer-function}
\end{equation}
where $\tilde{p}_w$ and $\tilde{v}_w$ are the temporal Fourier transforms of the time-domain wall-pressure $p^\prime_w$ and wall-normal velocity $v^\prime_w$ disturbance signals, respectively, and $\mathcal{H}(\omega)\in\mathbb{C}$ is a frequency-response transfer function with $\omega=2\pi f$. It is assumed that the PSub moves like a ``piano key" in the vertical direction only such that $\tilde{u}_w=\tilde{w}_w=0$. Disturbances are defined relative to the mean flow such that $p_w' = p - \overline{p}$ and $v_w' = v$. The transfer function $\mathcal{H}(\omega)$ governs the fluid–structure interaction at the wall and is engineered to yield a prescribed amplitude and phase response.  
The flow response is activated by a disturbance pressure field on the order of ten to a hundred Pascal and triggered by a surface phonon displacement on the order of a tenth of a micrometer. These response levels are compatible with MEMS-based material platforms at millimeter-to-centimeter scales~\cite{hopcroft2010young,zhou2018investigation}. The target phase can likewise be attained through mild structural dissipation~\cite{vignola2006effect} in the PSub components or, more broadly, through multiple-input/multiple-output configurations~\cite{willey2024tollmien}. 

For effective attenuation of boundary-layer instability waves, the wall response should scale as $\mathcal{O}\!\left(|\mathcal{H}(\omega)|\right)=1/\rho_e V_e$, where $\rho_e$ and $V_e$ denote the boundary-layer edge density and velocity, respectively. This scaling follows from the characteristic disturbance magnitudes since pressure fluctuations scale with the edge dynamic pressure, whereas velocity fluctuations scale with the edge velocity.

\begin{figure}[b]
\centering
\includegraphics[width=\linewidth]{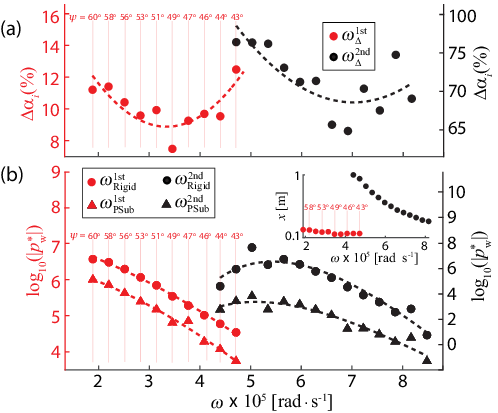}
\caption{Relative difference in maximum spatial amplification rate $\Delta \alpha_{i,max}$ (top) and N-factors $N=\log_{10}\left(p^*_w\right)$ at the end of the computational domain (bottom). All results are reported for the wave angles $\psi$, where the maximum amplification is observed.}
\label{fig:alpha-dns}
\end{figure}

The key insight of this work is that the phase relation between the wall-normal velocity and pressure fluctuations $\tilde{v}_w$ and $\tilde{p}_w$, denoted by
\begin{equation}
\Delta \phi_{p',v'} = \measuredangle\!\left(\tilde{p}_w,\tilde{v}_w\right),
\end{equation}
provides control authority over both first- and second-mode instability waves. While analogous transfer functions $\mathcal{H}(\omega)$ relating $\tilde{v}_w$ and $\tilde{p}_w$ can also be derived for porous surfaces and acoustic metasurfaces, they do not realize prescribed phase relations between these quantities across the relevant phase ranges; consequently, the resulting wall response does not permit control of the first-mode instability.

Effective suppression of both instability modes is demonstrated in Fig.~\ref{fig:alpha-dns} for a wide range of frequencies and the most amplified spanwise wavenumbers $\beta=2\pi/\lambda_z$ (or wave angles $\psi=\mathrm{atan}(\beta_r/\alpha_r)$). Figure~\ref{fig:alpha-dns}a shows the reduction in spatial growth rate derived from wall-pressure based amplitude curves,
\begin{equation}
\Delta \alpha_i = \frac{\alpha_{i,\mathrm{PSub}}-\alpha_{i,\mathrm{RW}}}{\alpha_{i,\mathrm{RW}}},
\end{equation}
evaluated at the streamwise locations where the maximum amplification occurs for the rigid-wall configuration. For the first mode, the maximum amplification occurs for oblique waves with $\beta \neq 0$ (or $\psi >0$)) as indicated, while the most amplified higher frequency second mode waves are 2-D ($\beta=0$). Overall, a significant reduction in growth rates is observed, which translates into substantial reductions in wall-pressure amplitudes and a delay in transition. Disturbance wall pressure amplitudes, $p_w^*=|\tilde{p}_w|/|\tilde{p}_{w,0}|$, were extracted at the end of the computational domain at $x=1.0$ m. The quantity $N\!\!=\!\!\log_{10}(p_w^*)$, commonly referred to as the $N$-factor, is shown for different frequencies, where $|\tilde{p}_{w,0}|$ denotes the disturbance amplitude at the first branch of the neutral curve$-$following standard practice in $N$-factor calculations. A clear reduction in disturbance amplitudes is observed across the entire frequency range for both first- and second-mode disturbances, demonstrating control effectiveness for monochromatic (single frequency tone) disturbance waves.

\begin{figure}[b]
\centering
\includegraphics[width=\linewidth]{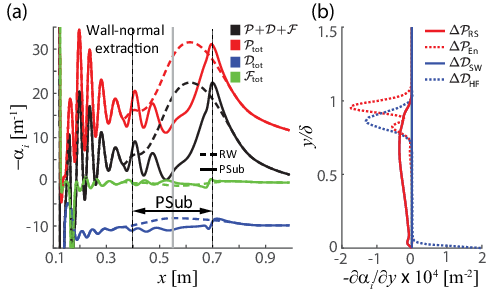}
\caption{Chu energy budget terms: production ($\mathcal{P}_{\mathrm{tot}}$), dissipation ($\mathcal{D}_{\mathrm{tot}}$), flux ($\mathcal{F}_{\mathrm{tot}}$), and their sum $(\mathcal{P}+\mathcal{D}+\mathcal{F})$ for rigid-wall and PSub configurations for a forcing frequency of $100 \mathrm{kHz}$. The reduction of the budget terms ($\Delta P$ and $\Delta D$) between controlled and uncontrolled cases is evaluated at $x=0.55$ m.}
\label{fig:budget-terms}
\end{figure}

The phase relations between disturbance variables are modified through the engineered interaction between the flow quantities at the wall and the PSub response, as further illustrated in Fig.~S2 of the Supplementary Material. The consequences of this phase manipulation are shown in Fig.~\ref{fig:budget-terms}, where the disturbance energy budget terms are displayed. The budget analysis is performed for a second-mode wave with a frequency of $100\,\mathrm{kHz}$ propagating in the Mach~5.35 boundary layer, with flow control applied between $x=0.4\,\mathrm{m}$ and $x=0.7\,\mathrm{m}$ that enforces a phase difference of $\Delta\phi_{p^\prime,v^\prime}=275^\circ$ at the wall. The different budget terms are normalized in a way that they can be interpreted as growth rates. The results show that local phase tuning leads to both a reduction of the total production term $\mathcal{P}_{\mathrm{tot}}$ and an increase in the total dissipation term $\mathcal{D}_{\mathrm{tot}}$. The dominant mechanism, however, is the suppression of the energy transfer from the mean flow to the disturbance flow field. This mechanism differs fundamentally from that of porous surfaces and metasurfaces, which attenuate second-mode disturbances primarily through increased dissipation and/or resonant scattering.\\
\begin{figure*}[t]
\centering
\includegraphics[width=0.75\textwidth]{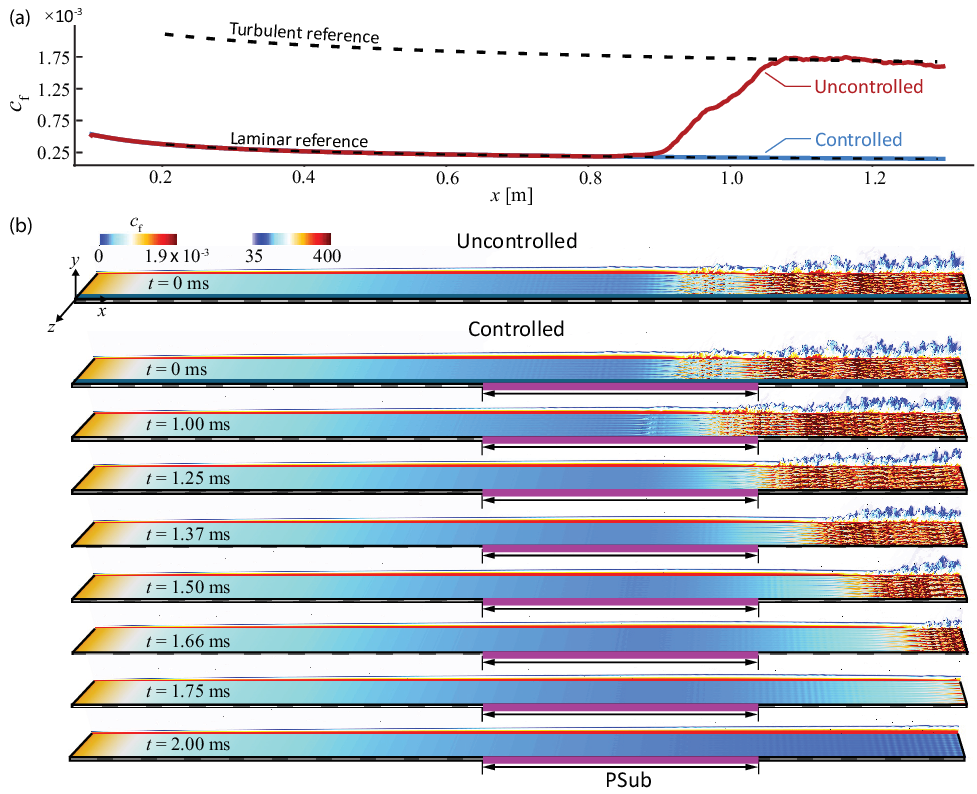}
\caption{PSub performance under broadband forcing simultaneously encompassing both first and second modes. (a) Time-averaged skin-friction coefficient for uncontrolled and PSub controlled cases. (b) Flow visualization of uncontrolled and PSub controlled DNS field. Temperature contours are shown on a cut plane in the background, and the wall surface is colored by skin-friction magnitude.}
\label{fig:breakdownDelay}
\end{figure*}
\indent Another notable observation is that the wall flux term $\mathcal{F}_{\mathrm{tot}}$, driven by the expression $\left< v^\prime p^\prime \right>$, although activated at the wall, does not play a dominant role in the disturbance energy budget$-$contrary to what one might intuitively expect. The differences in the production and dissipation terms throughout the boundary layer, denoted by $\Delta \mathcal{P}$ and $\Delta \mathcal{D}$, respectively, are shown in Fig.~\ref{fig:budget-terms} for the Reynolds stress production ($\Delta \mathcal{P}_{\mathrm{RS}}$), the entropy production ($\Delta \mathcal{P}_{\mathrm{En}}$), the shear work dissipation ($\Delta \mathcal{D}_{\mathrm{SW}}$), and the heat flux dissipation ($\Delta \mathcal{D}_{\mathrm{HF}}$) terms. Although the flow control approach interacts only directly with wall fluctuation quantities, its effect extends far into the boundary layer, even near the critical layer where both the entropy-production term and the heat-flux dissipation term peak. Moreover, the Reynolds-stress production term, which is active throughout the boundary layer, is reduced across the entire boundary layer. Similar observations are obtained for first-mode instabilities, despite the fact that they arise from fundamentally different physical mechanisms; the budget analysis results for these are shown in Fig. S3 in the Supplementary Material.\\
\indent One remaining question concerns the nature of the disturbance environment encountered in testing facilities and in free flight, where boundary-layer transition is typically triggered by broadband freestream disturbances. While the previous analysis focused on monochromatic forcing and thus examined the frequency response of individual instability waves, transition in realistic settings is driven by a broadband spectrum of instability waves. To demonstrate that the proposed flow-control design principle can be extended to broadband disturbances, we designed a transfer function $\mathcal{H}\left(\omega\right)$ as in Eq.~\ref{equ:transfer-function} that attenuates disturbances within the frequency–wavenumber spectrum $(f_m,\beta_n) \in [10,200]\,\mathrm{kHz} \times [0^\circ,170^\circ]$  discretized using $[m,n]=[0,20]\times[0,20]$ collocation points. The broadband disturbances are introduced through a volume-forcing term inside the boundary layer within the above frequency–wavenumber spectrum spanning the range of unstable first- and second-mode instability waves.

The flow conditions for the broadband case are identical to those considered previously, while the computational domain is extended to $x\in[0.1,1.3]$ m and $z\in[-0.025,0.025]$ m in the spanwise direction. In the baseline configuration, the flow undergoes transition to turbulence at approximately $x=1.0$ m, as indicated by the sharp rise in the skin-friction coefficient, which in the turbulent flow region closely follows the theoretical van~Driest turbulent skin-friction prediction~\cite{vanDriest1951turbulent}. Subsurface phonon intervention is applied within the region $x\in[0.5,1.0]$ m causing the boundary-layer flow to relaminarize over a time span of approximately 2 ms throughout the computational domain, with the skin-friction coefficient re-assuming the laminar skin friction values; see Fig.~\ref{fig:breakdownDelay}. This result underpins another key contribution$-$the effective control of first- and second-mode waves in a broadband disturbance environment.

In conclusion, this work introduces a new paradigm for passive control of hypersonic boundary-layer transition based on engineered phase relations between wall pressure and velocity fluctuations. By leveraging local phase and amplitude tuning by subsurface phonon engineering, simultaneous suppression of first- and second-mode instabilities is demonstrated to be physically possible, overcoming a long-standing limitation of passive laminar flow control. Energy budget analysis shows that the dominant stabilization mechanism arises from suppression of energy transfer from the mean flow to disturbance fluctuations through controlled phasing of disturbance quantities. DNS further demonstrates that this concept remains effective in broadband disturbance environments representative of realistic flight conditions, suggesting a viable pathway for extending laminar flow for full-scale hypersonic vehicles.
\\[1em]
\noindent \textbf{Acknowledgments} \\
\indent The authors are grateful to Dr. Kevin Bowcutt for fruitful discussions. This research is funded by the Office of Naval Research Multidisciplinary University Research Initiative (MURI) Grant No. N0001421268, with Dr. Eric Marineau and Dr. Jonathan Sosa serving as Program Managers. \\

\bibliographystyle{ieeetr}
\bibliography{FSIRefs}

\end{document}



\title{SUPPLEMENTARY MATERIAL
\\[1em]
Phonon-mediated stabilization of first and second modes \\in hypersonic boundary-layer flows}

\author{Christoph Brehm}
\email[]{cbrehm1@umd.edu}
\affiliation{Department of Aerospace Engineering, University of Maryland, College Park, Maryland 20740, USA}
\author{Connor W. Klauss}
\email[]{cwklauss@umd.edu}
\affiliation{Department of Aerospace Engineering, University of Maryland, College Park, Maryland 20740, USA}
\author{Mahmoud I. Hussein}
\email[]{mih@colorado.edu}
\affiliation{Smead Department of Aerospace Engineering Sciences, University of Colorado Boulder, Boulder, Colorado 80303, USA}
\affiliation{Department of Physics, University of Colorado Boulder, Boulder, Colorado 80302, USA}

\maketitle

This Supplementary Material provides additional background on the budget analysis results presented in the Letter, clarifies the general phase-tuning concept, and presents further results demonstrating the effectiveness of PSubs for boundary-layer control in the presence of broadband disturbances.

\section{Local Phase Tuning}\label{sectionS0}
\begin{figure*} 
\centering 
\includegraphics{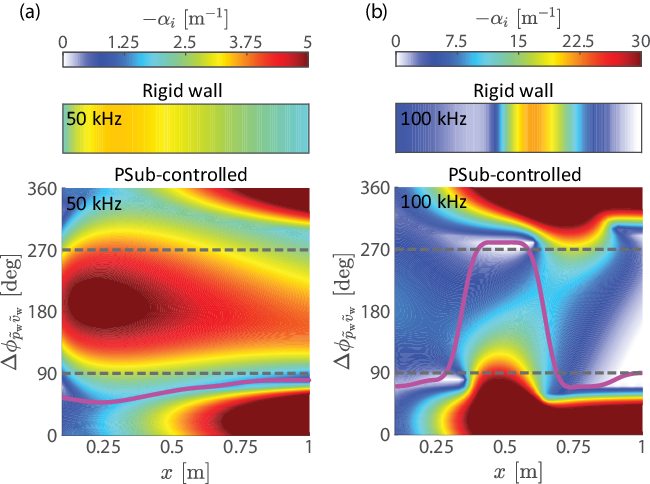}
\caption{Spatial amplification rates, $-\alpha_i$, from LST for the Mach 5.35 boundary-layer flow employing PSubs with an amplitude response of $\left|H_{\dot{y}}\right|=1.5$ and varying phase response $\Delta \phi_{p^\prime_w,v^\prime_w}$: (a) first mode at $f=50\,\mathrm{kHz}$ and (b) second mode at $f=100\,\mathrm{kHz}$.}
\label{fig:S0}
\end{figure*}
Throughout this Letter, we have emphasized that local phase tuning is the key principle through which phonon engineering can be used to stabilize boundary-layer instabilities, in particular the first and second modes. Figure~\ref{fig:S0} shows the leading unstable eigenvalue, $-\alpha_{i,\max}$, for two representative frequencies, $f=50\,\mathrm{kHz}$ and $f=100\,\mathrm{kHz}$, associated with the first and second modes, respectively. The stability results were obtained using the compressible linear-stability solver of Ref.~\cite{al2023biorthogonal}. At the wall, a PSub boundary condition was applied with an amplitude ratio of $\left|\mathcal{H}\left(\omega\right)\right|=2.5$, while the phase response was varied over $0^\circ \le \phi \le 360^\circ$. It should be noted that phase response in the range $90^\circ \le \Delta\phi_{p^\prime_w,v^\prime_w} \le 270^\circ$ can only be realized using a multiple-input multiple-output (MIMO) system, as discussed in Ref.~\cite{willey2023mimophononicsubsurfaces}, whereas the remaining phases can be realized using a single-input single-output (SISO) phononic subsurface (PSub)~\cite{hussein2015flowstab}, here represented by a single-degree-of-freedom damped harmonic oscillator. Moreover, a representative frequency-response function is shown in Section~\ref{sec::broadband-pulse}, where the phase shift is realized, with some approximation, without the use of damping in the system. The current results highlight the distinct stability characteristics of the first and second modes and their implications for the choice of the optimal phase $\Delta\phi_{p_w^\prime,v^\prime_w}$ required for effective boundary-layer control.

The largest amplification rates, corresponding to the most unstable conditions, occur for the first mode around $\Delta \phi_{p^\prime,v^\prime}\approx180^\circ$ for $x\lesssim 0.6~\mathrm{m}$ and around $\Delta \phi_{p^\prime,v^\prime}\approx30^\circ$ for $x\gtrsim 0.6~\mathrm{m}$. In contrast, for the second mode the largest amplification rates occur near $\Delta \phi_{p^\prime,v^\prime}\approx0^\circ$. For the first mode, a narrow low-growth band appears over $40^\circ \lesssim \Delta \phi_{p^\prime,v^\prime} \lesssim 90^\circ$, while stabilization of the second mode is obtained near $\Delta \phi_{p^\prime,v^\prime}\approx90^\circ$ and $\Delta \phi_{p^\prime,v^\prime}\approx270^\circ$. Both optimal phases for the first and second modes (solid line) fall into the SISO phase range, which for a monochromatic disturbance can be represented by a mass-spring-damper system matching the frequency response function (FRF) of a more complex PSub at a given frequency. These results emphasize that phonon engineering must take into account the underlying modal stability characteristics of the flow. When multiple instability modes coexist over a broad frequency band, the corresponding FRF must be tailored accordingly. For the results presented in this Letter, an optimization over the full spatial domain was carried out to determine an optimal phase, $\Delta \phi_{p^\prime,v^\prime}(f)$, that yields the desired stabilization for both monochromatic and broadband disturbances. It is also demonstrated in Section~\ref{sec::broadband-pulse} that this FRF can be realized considering prior work by the authors in Ref.~\citenum{hussein2015flowstab}.

\section{Chu Energy Budget Analysis}\label{sectionS1}

To elucidate the physical mechanisms underlying the stabilization of the first and second modes, we performed a disturbance energy budget analysis. The analysis is based on the Chu energy norm~\cite{chu1965energy},
\begin{equation}
  E=\frac{\overline{p}}{2\overline{\rho}^2}\tilde{\rho}\tilde{\rho}^* +\frac{\overline{\rho}}{2}\tilde{u}_i\tilde{u}_i^* + \frac{\overline{p}}{(\gamma-1)\overline{T}^2}\tilde{T}\tilde{T}^*
  ,
\end{equation}
where $\rho$, $u_i$, and $T$ denote the disturbance density, velocity components, and temperature, respectively, $\gamma$ is the specific heat ratio, overbars denote base-flow quantities, tildes denote complex temporal Fourier amplitudes, and $(\cdot)^*$ denotes complex conjugate quantities. 
The Chu energy norm is formally consistent only with temporal stability analysis (i.e., $\alpha \in \mathbb{R}$ and $\omega \in \mathbb{C}$). However, under the traveling-wave ansatz, the additional terms that arise in the spatial framework (i.e., $\alpha \in \mathbb{C}$ and $\omega \in \mathbb{R}$) have been shown to be negligible; see~\cite{al2023biorthogonal}. It can therefore be used to provide detailed insight into the relevant energy-transfer mechanisms. The individual production, dissipation, and flux terms can be written in the generalized form
\begin{equation}
\hat{\mathcal{T}} = -\int_{y=0}^\infty \Re\left< f^*\!\left(\boldsymbol{\tilde{q}}\right)
 h\!\left(\boldsymbol{\tilde{q}}\right)
\right> g\!\left(\overline{\boldsymbol{q}}\right) dy ,
\label{equ::gen-budget-term}
\end{equation}
where $\hat{\mathcal{T}}$ represent a generic budget term, such as production $\hat{\mathcal{P}}$, dissipation $\hat{\mathcal{D}}$, and flux $\hat{\mathcal{F}}$, $f$ and $h$ are complex functions of the disturbance flow state vector $\boldsymbol{\tilde{q}}$, $g$ is a real function of the base-flow state vector $\overline{\boldsymbol{q}}$, and $\Re[\cdot]$ denotes the real part.


Once the disturbance energy transport equation is derived, the individual contributions to disturbance-energy production, dissipation, and flux redistribution can be isolated. To enable a meaningful comparison between the controlled and uncontrolled cases, the budget terms are normalized to remove their dependence on disturbance amplitude, thereby isolating the regions responsible for energy production and dissipation. The resulting budget terms can be written as
\begin{equation} \begin{aligned} \alpha_{i,\mathrm{E}}(x,f) &= - \frac{ \hat{\mathcal{P}}_{\mathrm{tot}}+\hat{\mathcal{F}}_{\mathrm{tot}}+\hat{\mathcal{D}}_{\mathrm{tot}}} {\int_{0}^\infty 2\overline{u}E\,dy}\\ &= \underbrace{-\mathcal{P}_{\mathrm{tot}}-\mathcal{F}_{\mathrm{tot}}-\mathcal{D}_{\mathrm{tot}}} _{\alpha_{i,\mathcal{P}_{\mathrm{tot}}} + \alpha_{i,\mathcal{F}_{\mathrm{tot}}} + \alpha_{i,\mathcal{D}_{\mathrm{tot}}}} \end{aligned}, \label{eq::alpha_budgets} \end{equation}
where, after normalization, each term can be interpreted as a contribution to the spatial amplification rate with the integrand denoted by $\partial \alpha/\partial y$.~A reduction in disturbance-energy production and/or an increase in dissipation therefore translates directly into a reduction in modal growth rate. Following Ref.~\cite{montero2021analysis}, the total budget terms are decomposed as
\begin{equation} \begin{aligned} \mathcal{P}_{\mathrm{tot}}&=\mathcal{P}_{\mathrm{RS}}+\mathcal{P}_{\mathrm{Mom}}+\mathcal{P}_{\mathrm{En}}+\mathcal{P}_{\mathrm{PW}}+\mathcal{P}_{\mathrm{Dila}}+\mathcal{P}_{\mathrm{IE}}+\mathcal{P}_{\mathrm{TP}},\\ \mathcal{D}_{\mathrm{tot}}&=\mathcal{D}_{\mathrm{HF}}+\mathcal{D}_{\mathrm{SW}},~\mbox{and}\\ \mathcal{F}_{\mathrm{tot}}&=\mathcal{F}_{\mathrm{TP}}+\mathcal{F}_{\mathrm{HF}}+\mathcal{F}_{\mathrm{PW}}+\mathcal{F}_{\mathrm{SW}}. \end{aligned} \end{equation}
For the current Mach 5.35 flow case considered here, the dominant contributions are the Reynolds-stress and entropy-related production terms, $\mathcal{P}_{\mathrm{RS}}$ and $\mathcal{P}_{\mathrm{En}}$, together with the shear-work and heat-flux dissipation terms, $\mathcal{D}_{\mathrm{SW}}$ and $\mathcal{D}_{\mathrm{HF}}$. The remaining terms are at least an order of magnitude smaller and can, therefore, be neglected in the discussion below. 

To gain an in-depth understanding of the underlying mechanisms, we analyze the Reynolds stress production term. Following Eq.~\ref{equ::gen-budget-term}, with $f(\tilde{\boldsymbol{q}})=\tilde{u}_i$, $h(\tilde{\boldsymbol{q}})=\tilde{u}_j$ and the base-flow contribution given by $g(\overline{\boldsymbol{q}})=\partial \overline{u}_i/\partial x_j$, one obtains
\begin{equation}
\hat{\mathcal{P}}_{\mathrm{RS}}=-\int_{y=0}^{\infty}\overline{\rho}\Re\left<\tilde{u}_i\tilde{u}_j^*\right> \frac{\partial \overline{u}_i}{\partial x_j}dy,
\end{equation}
where $\partial \overline{u}_i / \partial x_j$ denotes the base-flow velocity gradients. This term depends on both the base-flow velocity gradients and the amplitude and phase relationships between the disturbance velocity components. The present control strategy is not intended to modify the base flow, which generally requires substantial energy input, for example through wall cooling/heating~\cite{bitter2015stability} and/or wall suction/blowing~\cite{zhuang2024instability}. Instead, it exploits the general structure of the Reynolds stress term
\begin{equation}
\Re\left< \tilde{u}_i  \tilde{u}^*_j \right> \sim |\tilde{u}_i||\tilde{u}_j|\cos\left(\Delta \phi_{u_i^\prime,u_j^\prime} \right),
\end{equation}
which depends on both the amplitudes of the disturbance velocities and their relative phase. The production term can therefore be reduced by appropriately tuning the phase differences between the velocity fluctuations, with analogous expressions applying to the remaining production terms. Because energy transfer from the mean flow to the disturbance field is governed by these terms, the present control approach suppresses modal amplification by manipulating phase relations among the disturbance quantities. More broadly, this phase-tuning concept may potentially also extend to other fields, such as electromagnetohydrodynamics, where instability control is of interest.


The complexity of phase manipulation becomes apparent when considering the detailed structure of the additional production and dissipation terms, for example,
\begin{equation}
\begin{aligned}
\mathcal{P}_{\mathrm{En}} &=
-\int_{0}^{\infty}
\overline{\rho}\,\mathcal{R}\,\gamma
\Re \left<\left(
\frac{\tilde{T}}{(\gamma-1)\overline{T}}
-
\frac{\tilde{\rho}}{\overline{\rho}}
\right)
\left(
\tilde{u}_{i}^{*}\frac{\partial \overline{T}}{\partial x_{i}}
\right)\right>\,dy,
\\
\mathcal{D}_{\mathrm{Visc}} &=
\int_{0}^{\infty}
\Re \left<\tilde{\tau}_{ij}
\frac{\partial \tilde{u}_{i}^{*}}{\partial x_{j}}\right>\,dy,
\\
\mathcal{D}_{\mathrm{HF}} &=
\int_{0}^{\infty}
\Re \left<\frac{\kappa}{\overline{T}}
\frac{\partial \tilde{T}}{\partial x_{i}}
\frac{\partial \tilde{T}^{*}}{\partial x_{i}}\right>\,dy,~\mbox{and}
\\
\mathcal{F}_{\mathrm{PW}} &=
\int_{0}^{\infty}\Re \left<
\frac{\partial \left(\tilde{v}\tilde{p}^{*}\right)}{\partial x_{i}}\right>\,dy.
\end{aligned}
\end{equation}
The strong coupling among the disturbance quantities renders phase manipulation inherently multidimensional and therefore nontrivial to optimize. This underscores that the simple notion of pure wave cancellation represents only a (lower) limiting case of the broader phase-tuning concept. In the present approach, both phase and amplitude modifications are imposed at the wall through the PSub. Given the complexity of the coupled disturbance dynamics, it was not evident \textit{a priori} that such wall-based tuning would result in substantial downstream flow stabilization.

\begin{figure*} 
\centering 
\includegraphics{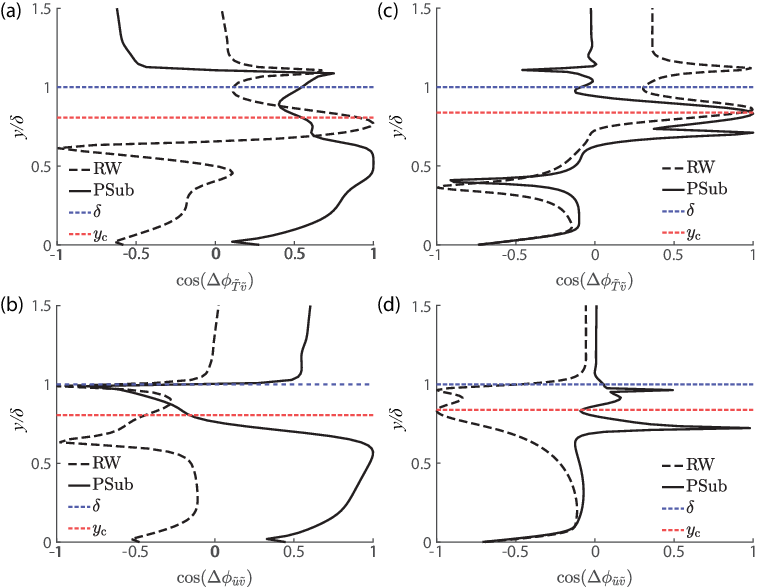}
\caption{Wall-normal distribution of the  direction cosine between temperature and wall-normal velocity fluctuations, $\cos\left(\Delta\phi_{\tilde{T}\tilde{v}} \right)$, for (a) $f=50\,\mathrm{kHz}$ and (b) $f=100\,\mathrm{kHz}$ and direction cosine between streamwise and wall-normal velocity fluctuations, $\cos\left(\Delta\phi_{\tilde{u}\tilde{v}} \right)$, for (c) $f=50\,\mathrm{kHz}$ and (d) $f=100\,\mathrm{kHz}$. Rigid-wall quantities are shown as dashed lines and PSub quantities are shown as solid.}
\label{fig:S1}
\end{figure*}
\begin{figure} [b]
\centering 
\includegraphics{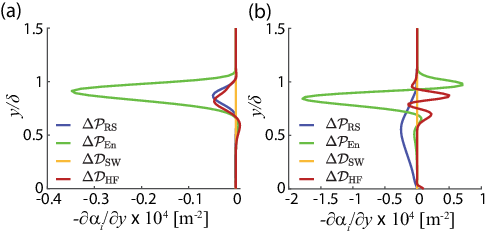}
\caption{Wall normal profiles of differences in dominant energy budget production/dissipation terms for the (a) 3D ($\beta=538$ m$^{-1}$) and (b) 2D (R, $\beta=0$ m$^{-1}$) 50 kHz first mode disturbance. For the 3D/2D modes,  profiles were extracted at $x=0.202$ m and $x=0.333$ m, respectively.}
\label{fig:S2}
\end{figure}

Figures~\ref{fig:S1}a--\ref{fig:S1}d show the distributions of the cosine of the phase differences between $T^\prime$ and $v^\prime$, and between $u^\prime$ and $v^\prime$, for the controlled (PSub) and uncontrolled (RW) cases at disturbance frequencies of 50 and 100 kHz. Solid lines denote the RW case, whereas dashed lines denote the PSub case. These variable pairs were selected because they dominate the relevant entropy-production, $\mathcal{P}_{\mathrm{En}}$, and Reynolds-stress-production, $\mathcal{P}_{\mathrm{RS}}$, terms. In particular, the leading contribution to the Reynolds-stress production is $\mathcal{P}_{\mathrm{RS}} \sim \overline{\rho}\Re\langle \tilde{u}_1 \tilde{u}_2^* \partial \overline{u}_1/\partial x_2 \rangle$, whereas the dominant entropy-production contribution is $\mathcal{P}_{\mathrm{En}} \sim \overline{\rho}\mathcal{R}\gamma\Re\langle \tilde{T}/[(\gamma-1)\overline{T}]\,\tilde{u}_2\,\partial \overline{T}/\partial x_2 \rangle$. Although the PSub interacts with the flow only at the wall, it alters the phase distribution throughout the boundary layer.

The corresponding effects on the relevant production and dissipation terms are shown in Figs.~\ref{fig:S2}a--\ref{fig:S2}c. For both first- and second-mode frequencies, the production terms are clearly reduced, whereas the dissipation terms change only weakly. This further emphasizes that the dominant stabilization mechanism is the disruption of energy transfer from the mean flow to the disturbance field through phase manipulation.

These observations are notable because the control does not primarily operate through enhanced damping or dissipation mechanisms, but through a redistribution of phase relations that weakens the dominant production pathways. In this sense, the PSub acts as a wall-based mediator of boundary-layer disturbance dynamics: a \textit{localized} boundary condition induces a distributed modification of the modal structure across the boundary layer. This provides direct physical support for the central premise of phonon-mediated stabilization.

\section{Broadband Pulse Simulation}\label{sec::broadband-pulse}

\begin{figure}[b] 
\includegraphics{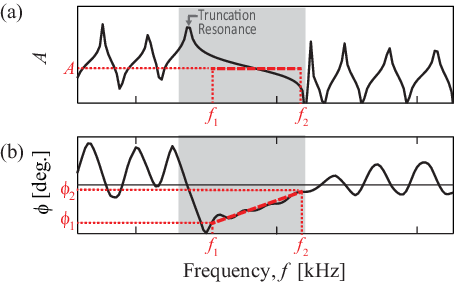}
\caption{Cartoon schematic of an arbitrary PSub frequency-response function (adapted from Ref.~\cite{hussein2015flowstab}). Superimposed is an approximate response function (red dashed lines) representaive of model used in the broadband simulations.}
\label{fig:S4}
\end{figure}

Finally, for a laminar-flow-control strategy to be viable in realistic settings, it must remain effective under broadband noisy disturbance forcing. Although Fig.~1 of the main article provides theoretical evidence of stabilization across the frequency ranges associated with the first and second modes for a hypothetical optimal phase response, $\Delta \phi^{opt}_{p^\prime,v^\prime}$, the broadband simulations require frequency-response functions that are realizable in the time-domain. Figure~\ref{fig:S4} shows a schematic of the the transfer function examined here, as an approximation to a realizable phonon-engineered response similar to the one featured in Ref.~\cite{hussein2015flowstab}. In the present example, a truncation resonance was introduced slightly below the relevant first- and second-mode frequency ranges to produce a target nearly constant-amplitude and linearly varying phase response, indicated by the red dashed lines over $f_1 \leq f \leq f_2$ and $\phi_1 \leq \phi \leq \phi_2$, respectively, where $f_1=10~\mathrm{kHz}$, $f_2=~200\mathrm{kHz}$, $\phi_1=0^\circ$, and $\phi_2=170^\circ$. Such a response provides a practically realizable approximation to the idealized phase tuning suggested by the monochromatic LST analysis, while remaining compatible with temporal simulations. The resulting flow response for this PSub configuration was analyzed using LST, and the results at $x=0.55\,\mathrm{m}$ are included in Fig.~1 of the main article. The stability characteristics reveal a compromise between stabilizing and destabilizing portions of the first- and second-mode frequency ranges, while still yielding a reduction in the overall peak amplification.

\begin{figure*}[t] 
\includegraphics{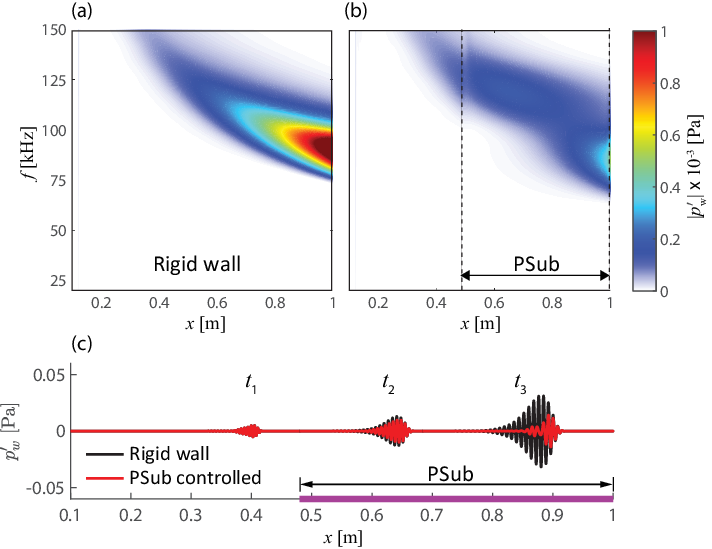}
\caption{Broadband stabilization of a pulse with a central forcing frequency of 100 kHz by a PSub with $\Delta\phi=180^\circ$ and $|H_{\dot{y}}=1.5|$ applied in $x\in[0.48, 1.0]$ m. Fourier-transformed wall pressure amplitude for rigid wall and PSub controlled cases are shown in (a) and (b). Time-instantaneous snapshots of disturbance pressure signals are given in (c).}
\label{fig:S3}
\end{figure*}

Prior to the large-scale DNS with broadband forcing presented as the final demonstration case in the main article, we employed the adaptive-mesh-refinement wavepacket-tracking (AMR-WPT) approach, an efficient framework for analyzing broadband disturbance responses developed in Refs.~\cite{browne2018_AMRWPT,browne2022nonlinear}. In this method, a pulse disturbance is introduced into the boundary layer, and the resulting spatio-temporally evolving wavepacket is tracked using AMR for computational efficiency. AMR-WPT provides an efficient means of characterizing the broadband stability properties of laminar boundary layers. Therefore, it serves as a natural intermediate step between the monochromatic LST analysis and the final large-scale broadband DNS. During the simulation, the disturbance field is monitored and processed to extract detailed information about the flow stability characteristics.

Figure~\ref{fig:S3} compares the Fourier-transformed wall-pressure signal for the rigid-wall and PSub cases. In the region where the PSub is applied, the disturbance level is markedly reduced. This reduction across a finite frequency band indicates that the control mechanism remains effective under broadband forcing and is not limited to a single tuned disturbance. Three representative snapshots of the wall-pressure disturbance at the wall are also shown in Fig.~\ref{fig:S3}. Together, these results show that the PSub remains an effective passive flow-control mechanism for delaying transition under broadband disturbances, thereby providing an important bridge between the monochromatic stability analysis and the final DNS demonstration. The reader is referred to Ref.~\cite{harris2025super} for an alternative approach for broadband PSub performance.

.

\bibliographystyle{ieeetr}
\bibliography{FSIRefs}